\documentclass[a4paper,twocolumn,pra,showpacs]{revtex4}
\usepackage{amsmath,amsfonts,amssymb}
\usepackage{amsfonts}
\usepackage{graphicx}
\usepackage{mathbbol}
\def\ra{\rangle}
\def\la{\langle}
\newcommand{\beq}{\begin{equation}}
\newcommand{\eeq}{\end{equation}}
\newcommand{\beqa}{\begin{eqnarray}}
\newcommand{\eeqa}{\end{eqnarray}}

\begin{document}

\title{Lewis-Riesenfeld invariants and transitionless tracking algorithm}

\author{Xi Chen$^{1,2}$}
\author{E. Torrontegui$^{1}$}
\author{J. G. Muga$^{1}$}
%
%\author{...}
%\thanks{}

\affiliation{$^{1}$Departamento de Qu\'{\i}mica-F\'{\i}sica,
UPV-EHU, Apdo 644, 48080 Bilbao, Spain}

\affiliation{$^{2}$Department of Physics, Shanghai University,
200444 Shanghai, China}


%\date{\today}

\begin{abstract}

Different methods have been recently put forward and implemented experimentally to inverse engineer the time dependent Hamiltonian of a quantum system and accelerate slow adiabatic processes via non-adiabatic shortcuts.
In the ``transitionless tracking algorithm'' proposed by Berry,  shortcut Hamiltonians are designed so that the system follows exactly, in an arbitrarily short time,
the approximate adiabatic path defined by a reference Hamiltonian. A different approach is based on designing first a Lewis-Riesenfeld invariant to carry the eigenstates
of a Hamiltonian from specified initial to final configurations, again
in an arbitrary time, and then constructing from the invariant
the transient Hamiltonian connecting these boundary configurations.
We show that the two approaches, apparently quite different in form and so far in results, are in fact strongly related and potentially equivalent, so that the inverse-engineering operations in one of them can be reinterpreted and understood in terms of the concepts and operations of the other one.
We study as explicit examples the expansions of time-dependent harmonic traps and state preparation of two level systems.
\end{abstract}

%\keywords{}

\maketitle
\section{Introduction}
%
%
%
%
%
%Slow adiabatic processes are quite common in quantum systems. In particular,
Externally imposed time-dependent interactions are frequently varied slowly to keep adiabaticity and control  the final state of a quantum system
robustly versus parameter fluctuations.
There are however many instances where we would like or need
to quicken these operations.
% duration of slow, adiabatic processes.
If they take too long, they are impractical for applications in which they are repeated many times, e.g. to manipulate
quantum information and transport ions or atoms \cite{Leibfried2002},
or they may suffer from decoherence, noise or losses, so
speeding them up may be the only way to actually implement
the ideal final outcome. Moreover in many experiments, as in atomic fountain clocks,
high repetition rates contribute to
achieve better signal-to-noise ratios and better accuracy \cite{Leanhardt}. Adiabatic steps are also the bottleneck in some cyclic processes, and determine, for example, the cooling rates in quantum refrigerators
quantifying the unattainability of absolute zero \cite{Salamon09,Kosloff-EPL,ChenPRA}.
%As a further motivation, let us mention the importance of adiabatic atomic transport.
%We would also like to preserve
%as much as possible the robustness characterizing the %adiabatic process.

Recently, several works have been devoted to theoretical proposals \cite{ion,MN08,Muga09,Ch10,Berry09,Salamon09,Calarco09,Ch10b,MN10,MN10b,Muga10,Li,transport} or experimental realizations
%For recent examples of this to implement interferometry with Bose-Einstein condensates see
\cite{David,S1,S2,Nice,Nice2} of fast non-adiabatic shortcuts to the states reached by a slow adiabatic processes in matter wave expansions or compressions, splitting, and transport.
%The adiabatic process can be accelerated with the ``fast-forward" scaling technique \cite{MN08,MN10,MN10b} and ``bang-bang" method \cite{Salamon09,Li}
%by using optimal control theory, which are extensively applied to
%some important issues on particle transport
%without vibrational heating \cite{David,ion,Calarco09,transport} and
%frictionless harmonic trap compressions or expansions
%for state preparation \cite{Salamon09,MN10,Rabitz98}.
Berry, in particular, has proposed a ``transitionless tracking algorithm'' to design time dependent interactions so that the system  follows exactly, in an arbitrarily short time,
the approximate adiabatic path defined by a reference, zeroth order Hamiltonian \cite{Berry09}. This method has been applied to
speed up adiabatic passage techniques and achieve fast and robust population control in two and three level atomic systems \cite{Ch10b}.
A different approach is based on designing first a Lewis-Riesenfeld invariant \cite{LR} to carry the eigenstates of a Hamiltonian from initial to final configurations, again
in an arbitrary time, and then constructing from the invariant
a transient, driving Hamiltonian \cite{Ch10,Muga09}. These methods were compared for harmonic oscillator expansions for which they provided rather different shortcut paths \cite{Muga10}.

Berry mentioned the existence of connections between the
invariants and the transitionless algorithm for a two-state system without pursuing them further \cite{Berry09}.
Following that hint we show in this work that these two approaches are in fact closely related, and can be stated in common terms, so that the inverse-engineering operations in one of them can be reinterpreted using concepts and operations of the other one. This sets their potential equivalence. The different results that have been found so far, as explained in detail below, are due to the ample freedom offered by both approaches to construct the driving shortcut Hamiltonian. We study the general setting as well as two explicit examples: expansions of time-dependent harmonic traps and state preparation in two level systems.

A word of caution on notation: as we shall deal with different methods and examples, multiple usage of some symbols such as
$|\phi_n \ra$, $|n\ra$, $H$, $I$, $\lambda_n$, $U$, or $E_n$ is unavoidable unless we load them with sub and superscripts, so consistency is strictly guaranteed only within each section.
In most cases this repeated usage will suggest a possible relation. %Indeed, connections among the two methods analyzed are set by %eventually interpreting the same symbols as the same object.
The context and explanations will clarify how this comes about.
\section{General framework}
\label{SecII}
\subsection{Lewis-Riesenfeld invariants}
We shall describe first Lewis-Riesenfeld theory in a
nutshell \cite{LR}.
Let us consider a quantum system evolving with a
time-dependent Hamiltonian $H(t)$.
A dynamical invariant $I(t)$ satisfies
\beq
i \hbar \frac{\partial I(t)}{\partial t } - [H(t), I(t)]=0,
\label{inva}
\eeq
so that its expectation values remain constant in time.
$I(t)$ can be used to express an arbitrary solution of the
time-dependent Schr\"{o}dinger equation
\beq
i \hbar \frac{\partial}{ \partial t } \Psi(t) = H(t) \Psi (t),
\label{tdse}
\eeq
as a superposition of ``dynamical modes'' $|\psi_n (t) \rangle$,
\beqa
|\Psi (t)\ra&=&\sum_n c_n  |\psi_n (t) \rangle,
\label{expan1}\\
|\psi_n(t)\ra&=&e^{i\alpha_n(t)}|\phi_n(t)\ra,
\label{expan2}
\eeqa
where $c_n$ are time-independent amplitudes, $|\phi_n (t)\rangle$ is the $n$-th eigenvector of the invariant $I(t)$, $I(t) |\phi_n(t) \rangle = \lambda_n |\phi_n(t) \rangle$, with $\lambda_n$ constant,
and the Lewis-Riesenfeld phase is defined as \cite{LR}
\beq
\label{LRphase}
\alpha_n (t) = \frac{1}{\hbar} \int_0^t \Big\langle \phi_n (t') \Big|
i \hbar \frac{\partial }{ \partial t'} - H(t') \Big| \phi_n (t')  \Big\rangle d t'.
\eeq
We use for simplicity a notation for a discrete spectrum of $I(t)$ but the generalization
to a continuum or mixed spectrum is straightforward. We also assume a non-degenerate spectrum.
\subsection{Invariant based inverse engineering}
Suppose that we want to drive the system
from an initial Hamiltonian $H(0)$, to a final one $H(t_f)$,
such that the populations in the initial and final instantaneous bases are the same, but admitting
transitions at intermediate times. To inverse engineer a time-dependent Hamiltonian $H(t)$ and achieve this goal, we
%proceed to use an inverse engineering approach
%to construct the time-dependent Hamiltonian $H$ by designing a time-dependent dynamics according to the invariant $I$.
%With the arbitrary eignstate and value, $|\phi_n (t) \rangle$ and $\lambda_n$, the dynamic invariant $I$
%can be expressed by
%
may define first the invariant through its eigenvalues and eigenvectors as
\beq
\label{invariant}
I(t) = \sum_n   |\phi_n (t) \rangle \lambda_n\langle \phi_n (t)|.
\eeq
The Lewis-Riesenfeld phases may also be chosen as arbitrary functions
to write down the
% dynamical mode $|\psi_n (t) \rangle$ of the yet-unknown time-dependent
%Hamiltonian $H(t)$ must satisfy Eq. (\ref{tdse})
%
%\beq
%i \hbar \frac{\partial}{\partial t} |\psi_n (t) \rangle = H (t)  %|\psi_n (t) \rangle,
%\eeq
%
%and has the form $e^{i \alpha_n (t)/\hbar} |\phi_n (t) \rangle $, as a dynamical mode of the invariant (\ref{invariant}), where $\alpha_n(t)$ can be set as an arbitrary time-dependent function.
%We can now construct the
time-dependent unitary evolution operator ${U}$, see Eqs. (\ref{expan1},\ref{expan2}), as
\beq
{U}= \sum_n e^{i \alpha_n (t)} |\phi_n (t) \rangle \langle \phi_n (0)|.
\eeq
It must obey
\beq
i \hbar \frac{\partial}{\partial t } {U}  = H(t) {U},
\eeq
which we solve formally for $H(t)$,
\beq\label{HfromU}
H(t)=i \hbar (\partial_t {U}){U}^{\dag},
\eeq
that is,
%
%\beqa
%H(t) = \sum_n  -\dot{\alpha}_n (t) |\phi_n (t) \rangle \langle \phi_n (t)|
%+ i \hbar | \partial_t \phi_n (t) \rangle \langle \phi_n (t)|.
%\eeqa
%
%or
%
\beq
\label{inHa}
H(t)=  F(t) + i \hbar \sum_n | \partial_t \phi_n (t) \rangle \langle \phi_n (t)|,
\eeq
where $F (t)$ is diagonal in the basis of the invariant,
\beq
F (t)=- \hbar \sum_n |\phi_n(t)\ra \dot{\alpha}_n  \la \phi_n(t)|,
\label{F}
\eeq
and the dot denotes derivative with respect to time.
Note that for a given invariant there are many possible
Hamiltonians corresponding to different choices of
phase functions $\alpha_n(t)$.\footnote{
%For later use and
To connect with
Lohe's work \cite{Lohe} we may also write
%(****??? EXPLAIN WHY, AT LEAST JUSTIFY MAYBE TO RELATE TO LOHE.
%WE DO NOT SEEM TO USE MUCH IF AT ALL THESE $T$ AND $I_0$
%LATER OR DO WE?) to define $I_{0}\equiv I(0)$
%
%\beq
%I_0=\sum_n |\phi_n (0) \rangle  \lambda_n  \langle\phi_n (0)|,
%\eeq
%
%
%\beq
$I(t)  =T I(0) T^{\dag},$
%\eeq
%
in terms of the unitary operator
%
%\beq
${T}= \sum_n |\phi_n (t) \rangle \langle \phi_n (0)|.
$
%\eeq
%
}
In general $I(0)$ does not commute with $H(0)$,
which means that the eigenstates of $I(0)$, $|\phi_n (0) \rangle$, do not coincide with the eigenestates of $H(0)$. $H(t_f)$ does not
necessarily commute with  $I(t_f)$ either.
If we impose $[I(0), H(0)]=0$ and $[I(t_f), H(t_f)]=0$,
the eigenstates coincide and
then a state transfer without final excitations is guaranteed.
%A simple, although by no means necessary relation is between initial and final invariants
%and Hamiltonians is
%$I(0)=a_0H(0)$ and $I(t_f)=a_fH(t_f)$, with $a_0$ and $a_f$ constants.
In a typical application the Hamiltonians $H(0)$ and $H(t_f)$ are given, they set the initial and final boundaries for the process, and we use these boundary conditions to define $I(t)$ and its eigenvectors accordingly. A convenient, although by no means necessary, relation is to set $I(0)=H(0)$.
We shall see specific examples of how this works
in Secs. III and IV.
\subsection{Transitionless tracking algorithm\label{tta}}
In Berry's method \cite{Berry09}, the starting point is a time-dependent reference Hamiltonian
\beq
H_0(t)=\sum_n | n_0(t)\rangle  E^{(0)}_n(t) \langle n_0 (t)|
\eeq
for which the approximate time-dependent adiabatic solutions
are
\beq
\label{aa}
|\psi_n (t) \rangle =  e^{i \xi_n (t)} |n_0(t)\rangle,
\eeq
where the adiabatic phases, with dynamical and geometric parts, are
\beq
\xi_n (t)=-\frac{1}{\hbar} \int^t_0 dt' E^{(0)}_n(t') +  i\int^t_0 dt' \langle n_0(t')| \partial_{t'} n_0(t') \rangle.
\eeq
The approximate adiabatic vectors in Eq. (\ref{aa}) are defined differently from the dynamical modes of the previous subsection, but they may potentially coincide, as discussed below, so, with some caution,
we use the same notation.
%so that
%
%
%\beq
%\label{solad}
%|\psi_n (t) \rangle =  \exp{\left\{- \frac{i}{\hbar} \int^t_0 dt' E_n(t') -  \int^t_0 dt' \langle n(t')| \partial_{t'} n(t') \rangle \right\}} |n(t)\rangle.
%\eeq
%
Defining now the unitary operator
\beqa
{U} &=& \sum_n e^{i \xi_n (t)} |n_0(t)\rangle \langle n_0(0)|,
\eeqa
a Hamiltonian $H(t)$ can be constructed, using again the general relation (\ref{HfromU}),
to drive the system exactly along the adiabatic paths of $H_0(t)$ as
%
%\beqa
%H = \nonumber \sum_n  (E_n- i \hbar \langle n | \partial_t n  \rangle )  |n \rangle \langle n |+ i \hbar \sum_n  | \partial_t n  \rangle \langle n|,
%\eeqa
%
%or
%
\beqa
\label{Berry Hamiltonian}
\nonumber
H(t)&=&  H_{0}(t) + H_1(t),
\\
H_1(t)&=& i \hbar \sum_n  \bigg(|\partial_t n_0(t)  \rangle \langle n_0(t) |
\nonumber
\\
&-&\langle n_0(t) | \partial_t n_0(t)  \rangle | n_0(t) \rangle \langle n_0(t) |\bigg),
\label{h1}
\eeqa
%
%where the explicit time-dependence of quantities has been suppressed for abbreviation, thus $| n \rangle$ represents $| n (t) \rangle $
where $H_1(t)$ is purely non-diagonal in the
$\{|n_0(t)\ra\}$ basis.

%where the time-dependent Hamiltonian $H^0$ with instantaneous eigenstates and energies is given by
%
%\beq
%H^0 (t) | n(t)\rangle = E_n(t) |n (t)\rangle.
%\eeq
%

We may change the functions $E^{(0)}_n(t)$, responsible for the dynamical part of the phase, and therefore $H_0(t)$ itself, keeping the same $|n_0(t)\ra$ eigenvectors. We could for example make all the $E^{(0)}_n(t)$  zero to suppress the dynamical phases, or compensate the geometric phase to have $\xi_n(t)=0$ \cite{Berry09}. Therefore, the Hamiltonian can be generally written in terms of the phases as
%
%\beqa
%\label{Generalized Berry Hamiltonian}
%H(t)= \sum_n \dot{g}_n |n(t)\rangle \langle n(t)|+ i \hbar \sum_n  | \partial_t n(t) \rangle \langle n(t)|,
%\eeqa
%
%which yields
%
\beqa
\label{NBH}
H (t)= G(t)+ i \hbar \sum_n  | \partial_t n_0(t)  \rangle \langle n_0(t)|,
\eeqa
where
\beq
G(t)=-\hbar \sum_n |n_0 (t) \ra \dot{\xi}_n  \la n_0 (t)|
\eeq
is diagonal in the instantaneous basis of $H_0$.
Subtracting $H_1$, $H_0$ may also be written as
\beq
\label{general H_0}
H_0 (t)=\sum_n |n_0 (t) \ra\big[i\hbar \la n_0(t)|\partial_t n_0 (t) \ra
- \hbar \dot{\xi}_n \big]\la n_0 (t)|.
\eeq
In general it is required that $H_1(t)$ vanish for $t<0$ and $t_f>0$, either suddenly or continuously at the extreme times. In that case the $|n_0 (t) \ra$ become also at the extreme times (at least at $t=0^{-}$ and $t=t_f^{+}$) eigenstates of the full Hamiltonian.

Using Eq. (\ref{inva}) and the orthonormality of the $\{|n_0(0)\ra\}$
it is easy to check that we may write invariants of $H(t)$ with the form
\beq
\label{inb}
I(t)=\sum_n |n_0(t)\ra \lambda_n\la n_0(t)|,
\eeq
where the $\lambda_n$ are constant eigenvalues.
For the simple choice $\lambda_n=E^{(0)}_n(0)$, then $I(0)=H_0(0)$.

Up to now we have presented the invariant-based and
tracking algorithm approaches in a common manner to make their relations obvious.
By reinterpreting the phases of Berry's method as $\xi_n (t)=\alpha_n (t)$, and the states as $|n_0(t)\ra=|\phi_n(t)\ra$, we may immediately equate $G (t) = F (t)$
and the Hamiltonians $H(t)$ in Eqs. (\ref{inHa}) and (\ref{NBH}).
%The approximate adiabatic
%modes of $H_0$ become exact dynamical modes of $H$ for the invariants
%(\ref{inb}).
We may also find the $H_0(t)$ implicit in the invariant's method using Eq. (\ref{general H_0}). In other words, the dynamical modes in the invariant-based method can be also understood as approximate adiabatic modes of a certain Hamiltonian $H_0(t)$.

An important caveat is that, although the two methods could coincide, they do not have to. Given $H(0)$ and $H(t_f)$ there is much freedom to connect them using different invariants, phase functions, and reference Hamiltonians $H_0(t)$. It should be clear by now that each of these methods does not provide a unique shortcut but entire families of them, a welcome flexibility that allows to optimize the path according to physical criteria and/or or operational constraints.

%**??**So far, we have clarified the relation between Berry's general transitionless tracking algorithm and inverse engineering method
%in the same frame of Lewis-Riesenfeld invariant theory. However, it seems that in the applications of shortcut to adiabaticity, the invariant-based inverse engineering method
%and transitionless tracking algorithm are quite different in form. In detail,
%the Hamiltonian (\ref{inHa}) will drive its eigenstate from initial to final one,
%with appropriate boundary conditions, along the invariant dynamics. Whereas, the Hamiltonian (\ref{Berry Hamiltonian}) including the fast-driving term $H_1 (t)$
%always drives the state along the adiabatic passage of $H_{0}$ as reference.
In the following sections, we shall work out two specific examples
where the connections, differences and similarities of the
two approaches are illustrated and examined further.
\section{Time-dependent harmonic oscillator}
\subsection{Lewis-Riesenfeld invariants}
The Hamiltonian of a particle in a time-dependent harmonic oscillator with varying
angular frequency (as all ``frequencies'' hereafter) $\omega(t)$, is
\beq
\label{effective ho}
H (t)=\frac{1}{2 m} \hat{p}^2+ \frac{m \omega^2(t)}{2}  \hat{q}^2.
\eeq
The instantaneous eigenstates and energies are, respectively,
\beqa
\label{is-ho}
\la x | n (t) \ra &=&\left(\frac{m\omega(t)}{\pi\hbar}\right)^{1/4}
\!\frac{1}{(2^n n!)^{1/2}}
\nonumber\\
&\times& \exp{\left[-\frac{m \omega(t)}{2\hbar}x^2\right]}
H_n\left[\sqrt{\frac{m\omega(t)}{\hbar}}{x}\right],
\eeqa
where $H_n$ is a Hermite polynomial,
and $E_n (t) = (n+1/2) \hbar \omega(t)$,
so that $H(t) |n (t)\rangle= E_n(t) |n (t)\rangle$.
%According to the Lewis-Riesenfeld invariant theory \cite{LR},
%the exact solution of the time-dependent Schr\"{o}dinger equation,
%can be written as the superposition of orthogonal ``expanding modes''
%$\psi_n$, as in Eqs. (\ref{expan1},\ref{expan2}).
%
%\beq
%\Psi(t,x)=\sum_n c_n e^{i \alpha_n(t)/\hbar} | \phi_n (t) \rangle,
%\eeq
%
Trying a quadratic ansatz for the invariant \cite{LR,Lohe,Berry,Dodonov},
\beq
I (t)= \frac{1}{2} \left[(1/b^2) \hat{q}^2m \omega_0^2+\frac{1}{m}
\hat{\pi}^2 \right],
\eeq
where $\hat{\pi}=b(t)\hat{p}-m\dot{b}\hat{q}$ plays the role of a momentum conjugate to $\hat{q}/b$, and inserting it into Eq. (\ref{inva}),
the scaling factor $b=b(t)$ is found to satisfy the Ermakov equation \cite{LR,Lohe}
\beq
\label{ermakov}
\ddot{b}+\omega^2(t) {b}=\frac{\omega_0^2}{b^3}.
\eeq
$\omega_0$ is in principle an arbitrary constant, which we fix as the initial frequency.
The eigenfunctions of $I(t)$ are
\beqa
\label{emode}
\la x | \phi_n (t) \ra &=&% \left(\frac{m\omega_0}{\pi\hbar}\right)^{1/4}
\!\frac{1}{(2^n n! b)^{1/2}} \exp{\left[ i \frac{m}{2\hbar}\left(\frac{\dot{b}}{b} +
 \frac{i\omega_0}{b^2}\right)x^2 \right]}
\nonumber\\
&& \times
H_n \left[\left(\frac{m\omega_0}{\hbar}\right)^{1/2}\frac{x}{b}\right],
\eeqa
and, since $I(t)$ has the structure of a generalized harmonic oscillator,  i.e., quadratic but with ``crossed'' momentum-position terms,
the eigenvalues are
$\lambda_n=(n +1/2) \hbar \omega_0$.
%According to Eq. (\ref{invariant}), the invariant can thus be constructed by
%
%\beq
%I (t) = \left(n +\frac{1}{2}\right) \hbar \omega_0  | \phi_n (t) \rangle \langle \phi_n (t) |,
%\eeq
%
%which can be explicitly expressed by
%
%\beq
%\label{invariant-ho}
%I (t) = \frac{1}{2m} {(b \hat{p})}^2 + \frac{m}{2}  \left(\dot{b}^2 + \frac{\omega^2_0}{b^2}\right) \hat{q}^2 - \frac{b \dot{b}}{2} (\hat{p} \hat{q} +\hat{q} \hat{p}),
%\eeq
%

Substituting the wave function $\la x | \phi_n(t) \ra$ into Eq. (\ref{inHa}), the Hamiltonian may be written as
\beq
H (t)= F (t) + \frac{\dot{b}}{2 b} (\hat{p} \hat{q} +\hat{q} \hat{p})-  \frac{m}{2}  \left(\dot{b}^2 + \frac{\omega^2_0}{b^2}\right) \hat{q}^2,
\eeq
For consistency with Eq. (\ref{effective ho}) the crossed terms must cancel. This is indeed the case.
$F (t)$ can be determined, using Eq. (\ref{F}), from the phase of the dynamical modes,
Eq. (\ref{LRphase}), which is now
%Since $[I,  i \hbar \frac{\partial }{ \partial t} - H] =0$,
%we can choose  the Lewis-Riesenfeld phase, defined by Eq. (\ref{A-A phase}), in $\exp{[- i f_n (t)/\hbar]}$. In this case,
%
\beq
\alpha_n (t)
% &=& \nonumber  -\int_0^t  \Big{\langle} \phi_n (t') \Big| H(t') - i \hbar \frac{\partial  } {\partial t'}\Big| \phi_n (t') \Big{\rangle} dt',
%\nonumber\\
=  -\left(n+\frac{1}{2}\right) \omega_0  \int_0^t \frac{1}{b^2} dt'.
\eeq
This gives $\dot{\alpha}_n = -(n + \frac{1}{2}) \omega_0/ b^2$ and $F(t)= I(t)/b^2$. The Hamiltonian $H(t)$ can be finally written as
\beq
\label{hohb}
H (t) = \frac{1}{2m} {\hat{p}}^2 + \frac{1}{2} m \left( \frac{\omega^2_0}{b^4}-\frac{\ddot{b}^2}{b} \right) \hat{q}^2,
\eeq
which is nothing but the Hamiltonian (\ref{effective ho}) after substituting $\omega(t)$ using the Ermakov
equation (\ref{ermakov}).
%As mentioned above, this Hamiltonian drives the state along the expanding modes,
%$|\psi_n (t) \ra = e^{i \alpha_n(t)} | \phi_n (t) \rangle$, with %$\alpha_n(t)=-(n+1/2)\omega_0\int_0^t dt'/b^2$, as descried in Ref. \cite{Ch10}.
%
%
%
%
\subsection{Invariant-based engineering approach
\label{ibea}}
In the inverse engineering approach based on Lewis-Riesefeld invariant theory
as presented in \cite{Ch10}, the main goal is to find a ``trajectory'' for the external parameter $\omega(t)$
so that the populations of the final oscillator levels are the same as the populations of the initial one. Designing $I(t)$ first here means to design $b(t)$.
%Comparison between Eqs. (\ref{Hamiltonian-harmonic oscillator}) and (\ref{effecive ho}) suggests that
%the well-known Ermakov equation can be applied to design time-dependent frequency $\omega(t)$ of harmonic oscillator by the scaling parameter $b(t)$.
%As discussed in Sec. \ref{SecII}, $H(t)$ and $I(t)$ do not necessarily commute, so the
%expanding modes $|\psi_n(t) \ra$ are in general not the same as the instantaneous eigenstates (\ref{emode}),
%and may have more than one component in the ``adiabatic basis'' of instantaneous eigenstates (\ref{is-ho}).
Let us assume an expansion with initial and final frequencies $\omega (0)= \omega_0$ and $\omega (t_f)= \omega_f$.
To make $I(t)$ and $H(t)$ commute at $t=0$ and $t_f$
and have common eigenfunctions we impose the boundary conditions
\beqa
b(0) &=& 1, ~~\dot{b} (0)=0,~~ \ddot{b} (0)=0,
\nonumber
\\
b(t_f)&=&\sqrt{\omega_f/\omega_0}, ~\dot{b} (t_f)=0,~ \ddot{b} (t_f)=0,
\label{bc}
\eeqa
by comparing Eqs. (\ref{is-ho}) and (\ref{emode}) and using Eq. (\ref{ermakov}).
%have to be imposed to make $I(0)$ and $I(t_f)$ commute with $H(0)$ and $H(t_f)$.
%Any other choice would necessarily produce ``frictional heating''.
A simple polynomial ansatz can be used to interpolate $b(t)$
at intermediate times \cite{Ch10}.
%\beq\label{ans}
%$b(t)=\sum_{j=0}^5 a_j t^j$,
%\eeq
%
%can be used \cite{Ch10}.
%solved to provide the coefficients
%
%\beq
%b (t) =
%6 \left(\gamma -1\right) s^5
%-15 \left(\gamma-1\right) s^4 +10 \left(\gamma-1\right)s^3
%+ 1,
%\eeq
%
%where $s=t/t_f$.
Once $b(t)$, and therefore the invariant, are set, the time-dependent frequency $\omega(t)$ and the shortcut Hamiltonian follow from the Ermakov equation (\ref{ermakov}).
This method, including the effect of gravity, has been realized experimentally \cite{Nice}, and extended to Bose-Einstein condensates \cite{Muga09,Nice2}, and to
design transport protocols \cite{transport}.
%Thus, by engineering $\omega(t)$, we can
%drive from the initial eigenstates of the Hamilton $H(t)$ to final one along the invariant dynamics without any transition at time $t=0$ and $t=t_f$, although
%in some time interval the harmonic trap may become an expulsive parabolic potential for much shorter time $t_f$,
%and the transient energy excitation could be very high in the intermediate time \cite{ChenPRA}.
%
%
%
%
\subsection{Invariant-based method in transitionless-tracking-algorithm language}
%
%
%
%
%When Berry's transitionless algorithm is applied,
%the central idea is to construct a Hamiltonian $H(t)$ that drives the system exactly along the adiabatic paths
%defined for the time-dependent Hamiltonian $H_0 (t)$ as reference, %without any transitions among the instantaneous
%eigenstates of $H_0 (t)$.
As pointed out in Sec. \ref{tta}, the invariant based method can be restated in the language of the transitionless tracking algorithm.
To find the reference Hamiltonian $H_0(t)$ implicit in Sec. \ref{ibea},
we interpret the $|\phi_n (t)\ra$ as the eigenstates $|n_0(t)\ra$ of $H_0$, take $\xi_n (t)=\alpha_n(t)$, and set the eigenvalues according to
Eq. (\ref{general H_0}). This gives
%
%\beqa
%\alpha_n (t) = -\int_0^t  \Big{\langle} \phi_n (t') \Big| H_0(t') - i \hbar \frac{\partial} {\partial t'}\Big| \phi_n (t') \Big{\rangle} dt',
%\eeqa
%
%so that the instantaneous eigenvalues of Hamiltonian $H_0 (t)$, satisfying $H_0 (t) |\phi_n (t) \rangle = \varepsilon_n | \phi_n (t)  \rangle$, can be expressed by
%
%\beq
%\varepsilon_n = -\dot{\alpha_n} (t) + i \hbar \langle \phi_n (t) | \partial_ t \phi_n (t)  \rangle.
%\eeq
%
%We choose as before $\dot{\alpha_n} (t)= -(n+1/2) \hbar \omega_0/b^2$.
%Setting $\dot{g_n} =- \dot{\alpha_n}$ and $|n\ra = |\phi_n \ra$,
%then
%Substituting into Eq. (\ref{general H_0}), we obtain
%
\beqa
H_0 (t) =
\left( \frac{1}{b^2} + \frac{\dot{b}^2 - \ddot{b} b }{2 \omega^2_0} \right)I(t),
\eeqa
which is a generalized harmonic oscillator.
The corresponding Hamiltonian $H_1 (t)$ takes the form
\beq
H_1 (t) =  \frac{\dot{b}}{2 b} (\hat{p} \hat{q} +\hat{q} \hat{p}) -  \frac{m}{2}  \left(\frac{\ddot{b}^2}{b} + \frac{\dot{b}^2}{b^2}\right) \hat{q}^2 -\left(\frac{\dot{b}^2 - \ddot{b} b }{2 \omega^2_0} \right) I.
\eeq
Because of the boundary conditions (\ref{bc}),
$H_1 (t)$ vanishes at $t=0$ and $t=t_f$.
The crossed terms are canceled out in the
full Hamiltonian $H(t) = H_0(t) + H_1(t)$ given by Eq.
%
%\beq
%\label{exact}
%H (t)  = \frac{1}{2m} {\hat{p}}^2 + \frac{1}{2} m \left( \frac{\omega^2_0}{b^4}-\frac{\ddot{b}^2}{b} \right) \hat{q}^2,
%\eeq
%
(\ref{hohb}).
%The solution, $ |\psi_n (t) \ra = e^{ i \alpha_n(t)/\hbar} | \phi_n (t) \rangle $,
%being the solution of Hamiltonian $H_0 (t)$ in the adiabatic approximation, becomes the exact dynamics of the Hamiltonian
%$H(t)$, where the Hamiltonian $H_0 (t)$ as reference can be considered as
%a generalized harmonic oscillator \cite{Berry}. So we
%achieve the state dynamics without transitions among the instantaneous eigenstates of Hamiltonian $H_0 (t)$ at any time.
%However, the physical realization of $H_1 (t)$ in the intermediate time will become %problematic due to the nonlocal interaction \cite{Muga10}.
%
%
%
\subsection{Transitionless tracking algorithm (standard application)}
Unlike the previous subsection, in a more standard application
of the transitionless tracking algorithm
to the harmonic oscillator \cite{Muga10},
$H_0(t)$ is set first as an
ordinary harmonic oscillator with given frequency $\omega(t)$, i.e.,
it takes the form (\ref{effective ho}), so that $|n (t) \ra$ in Eq. (\ref{is-ho}) should be now reinterpreted as $|n_0 (t) \ra$. $H_1(t)$
is calculated from Eq. (\ref{h1}), and the resulting shortcut Hamiltonian becomes
%
%As a matter of fact, in Berry's transitionless algorithm, the general Hamiltonian can be constructed by
%Eq. (\ref{New Berry Hamiltonian}). In a particular case, using Eq. (\ref{solad}),
%we can calculate the Hamiltonian $H(t)$ from Eq. (\ref{Berry Hamiltonian}) and obtain
%
\beqa
\label{Berry Hamiltonian-ho}
H (t)=\underbrace{ \frac{{\hat{p}}^2}{2m}  + \frac{1}{2} m  \omega^2 (t) \hat{q}^2}_{H_{0} (t)} \underbrace{- \frac{\dot{\omega}}{4 \omega (t)} (\hat{p} \hat{q} +\hat{q} \hat{p})}_{H_1 (t)},
\eeqa
%
%where $H_{0} (t)$ is equal to the effective Hamiltonian (\ref{effecive ho}). %, satisfying  $H_{0} (t) | n(t) \rangle =E_n (t) | n(t) \rangle$
%and $E_n(t) = (n+1/2) \hbar \omega(t)$.
a generalized harmonic oscillator
with crossed terms that imply a non-local interaction \cite{Berry}.
%Since the states in the adiabatic approximation driven by
%$H_0 (t)$ become the exact dynamics of the Hamiltonian $H (t)$,
%the whole Hamiltonian $H (t)$ with the help of $H_1 (t)$ will drive
%the states without any transitions with respect to $H_{0} (t)$.
%So this approach provides an alternative way to achieve the shortcut to adiabaticity.
%Nevertheless, for the time-dependent harmonic oscillator, the Hamiltonian $H(t)$, can be considered as a generalized
%harmonic oscillator \cite{Berry}, and becomes problematic to implement experimentally for expansion \cite{Muga10}  and transport of harmonic trap \cite{transport},
%due to the nonlocal operator $H_1 (t)$.
%
%
%
%
\subsection{Relation to invariants}
To reinterpret the previous subsection in terms of invariants, we construct $I (t)= \sum_n | n_0(t)\rangle \lambda_n \langle n_0(t)|$, with $\lambda_n = E_n^{(0)}(0)=(n +1/2) \hbar \omega_0$.
Since $H_0 (t)= \sum_n | n_0(t)\rangle E_n^{(0)}(t)\langle n_0 (t)|$ with instantaneous eigenvalues $E_n^{(0)}(t)=  (n +1/2) \hbar \omega (t) $,
the invariant is now proportional to an ordinary harmonic oscillator,
\beq
I (t) = \frac{\omega_0}{\omega(t)} H_0 (t) = \frac{\omega_0}{\omega(t)} \left[ \frac{{\hat{p}}^2}{2m}  + \frac{1}{2} m  \omega^2 (t) \hat{q}^2 \right].
\eeq
and $I(0)=H_0(0)$.
Using $\dot{\xi}_n = -E_n^{(0)}(t)/\hbar + i \la n_0(t)| \partial_t n_0(t) \ra$, and $\la n_0(t)| \partial_t n_0(t) \ra =0$,
then letting $\alpha_n (t) = \xi_n (t)$ and $ |\phi_n (t) \ra =|n_0 (t)\ra$, we
may write down $H(t)$ from Eq. (\ref{inHa}),
\beqa
H (t)= F(t) - \frac{\dot{\omega}}{4 \omega (t)} (\hat{p} \hat{q} +\hat{q} \hat{p}),
\eeqa
but here $F(t)=H_0(t)=[\omega(t)/\omega_0] I(t)$, so we recover the Hamiltonian (\ref{Berry Hamiltonian-ho}).
%
%\beq
%H (t)=\frac{{\hat{p}}^2}{2m}  + \frac{1}{2} m  \omega^2 (t) \hat{q}^2 - \frac{\dot{\omega}}{4 \omega (t)} (\hat{p} \hat{q} +\hat{q} \hat{p}),
%\eeq
%
%with $F (I,t)= $. The above Hamiltonian is the same as the Hamiltonian (\ref{Berry Hamiltonian-ho}) obtained by Berry's method.
%It is seen that Berry's transitionless tracking algorithm is different from the invariant-based inverse engineering approach, at least in form.
%The invariant $I(t)$ obtained here has the form of effective harmonic oscillator, rather than, a generalized harmonic oscillator.
$I(t)$ does not commute with $H(t)$ in general. To
guarantee that $I (t)$ and $H (t)$ have common eigenstates at $t=0$ and $t=t_f$, the boundary conditions $\dot{\omega}(0)=0$ and $\dot{\omega}(t_f)=0$ should be satisfied, so that $H_1 (t)$ vanishes at the initial and final times, as discussed in Sec. \ref{SecII}.
\section{Two-level atom}
The two-level atom is another fundamental model.
Speeded-up adiabatic state preparation methods such as Rapid Adiabatic Passage (RAP) in two-level atomic system may be useful in chemical reaction dynamics, laser cooling, or quantum information processing.

For the two-level atom, using $|1\ra={0\choose 1}$, $|2\ra={1\choose 0}$, the time-dependent Hamiltonian which we consider, in a laser adapted interaction picture
and applying the rotating wave approximation, is
\beqa
\label{H-2level}
H (t)= \frac{\hbar}{2} \left(\begin{array}{cc} \Delta &\Omega_{R} e^{i \varphi}
\\
\Omega_{R} e^{-i \varphi}&-\Delta
\end{array}\right),
\eeqa
where $\Delta=\Delta(t)$ and $\Omega_{R}=\Omega_R(t)$ are the time-dependent detuning and Rabi frequency, and $\varphi=\varphi(t)$ a time-dependent phase.
The instantaneous eigenvectors are
\beqa
\label{instantaneous states-1}
|n_{+}(t)\rangle &=& \cos{\left(\frac{\theta}{2}\right)} e^{i \varphi} |2 \rangle + \sin{\left(\frac{\theta}{2}\right)} |1 \rangle,
\\
\label{instantaneous states-2}
|n_{-}(t)\rangle &=&  \sin{\left(\frac{\theta}{2}\right)}  |2 \rangle - \cos{\left(\frac{\theta}{2}\right)} e^{-i \varphi} |1 \rangle,
\eeqa
with the mixing angle $\theta=\theta(t)\equiv \arccos (\Delta/\Omega)$
and  eigenvalues $E_{\pm}(t) = \pm \hbar \Omega (t) /2$, where $\Omega (t)= \sqrt{\Delta^2 + \Omega_R^2}$.
If $\varphi=0$, and the adiabaticity condition
\beqa
\left|\frac{\Omega_R \dot{\Delta}-\dot{\Omega}_R \Delta}{\Omega^3}\right|\ll 1,
\eeqa
is satisfied, the state evolving from $|\psi_{\pm} (0) \rangle= |n_\pm(0)\rangle$ follows the adiabatic approximation
\beqa
\label{adiabatic states-two level atom}
|\psi_{\pm}(t)\rangle= \exp{\left\{-\frac{i}{\hbar}\int^{t}_{0} dt' E_\pm (t')\right\}} |n_\pm(t)\rangle,
\eeqa
whereas transitions will occur otherwise. If the state starts from  $|n_{+}(0)\rangle$,
the adiabatic evolution of the population of levels $1$ and $2$ is
\beqa
P^{ad}_1(t)&=& |\la 1 |n_{+}(t)\rangle |^2 = \sin^{2}{\left(\frac{\theta}{2}\right)},
\nonumber\\
P^{ad}_2(t)&=& |\la 2 |n_{+}(t)\rangle |^2  = \cos^{2}{\left(\frac{\theta}{2}\right)}.
\label{pad}
\eeqa
In what follows we shall speed up the adiabatic passage by the two methods presented here.
\subsection{Invariants method}
\label{SecIVA}
To use the invariant-based inverse engineering method, we first
parameterize the eigenvalues and eigenstates of
the invariant $I(t)$, satisfying $I(t) |\phi_n (t)\rangle = \lambda_n |\phi_n (t)\rangle$ consistently with orthogonality and normalization, in parallel to Eqs. (\ref{instantaneous states-1}) and (\ref{instantaneous states-2}),
\beqa
\label{instantaneous states I-two level}
|\phi_{+} (t) \rangle &=&   \cos{\left(\frac{\gamma}{2}\right)} e^{i \beta} |2 \rangle + \sin{\left(\frac{\gamma}{2}\right)}  |1 \rangle,
\\
|\phi_{-} (t) \rangle &=& \sin{\left(\frac{\gamma}{2}\right)}  |2 \rangle  -  \cos{\left(\frac{\gamma}{2}\right)} e^{-i \beta}
|1 \rangle,
\eeqa
and set $\lambda_\pm =  \pm \hbar\Omega_0 /2$. Thus, $I(t)$ can be expressed as
\beqa
\label{I}
I (t)= \frac{\hbar}{2}\Omega_0 \left(\begin{array}{cc} \cos\gamma & \sin{\gamma} e^{ i \beta}
\\ \sin{\gamma} e^{- i \beta} &  -\cos \gamma
\end{array}\right),
\eeqa
where $\beta=\beta(t)$ and $\gamma=\gamma(t)$ are auxiliary time-dependent angles.
%Physically, it is required that the initial state $|\phi_n (0) \rangle$
%be a pure ground state $|1 \rangle$ or excited state $|2 \rangle$,
%so that we choose $I_0=I(0)$, $\Omega_0=\Omega(0)$ and
%%
%\beq
%I_0 = \frac{\hbar}{2}\Omega_0 \hat{\sigma}_z.
%\eeq
%
%and
%$I (t) = \hat{T}^{\dag} I_0 \hat{T}$ where the time-dependent unitary evolution operator is \cite{Lai},
%
%\beq
%\hat{T} = \exp{\left[\frac{\gamma (t)}{2} ( \hat{\sigma}_{+} e^{
%i \beta(t)} + \hat{\sigma}_{-} e^{- i \beta(t)})\right]}.
%\eeq
%
%As pointed out in \ref{SecII}A, the solution time-dependent Schr\"{o}dinger equation
%$i \hbar (\partial/\partial t) |\psi_{\pm} (t) \rangle = H (t)  |\psi_{\pm} (t) \rangle$
%can be written as
%$|\psi_{\pm} (t) \rangle= e^{ i \alpha_{\pm} (t)} | \phi_{\pm} (t) \rangle$,
Using Eqs. (\ref{H-2level}) and (\ref{instantaneous states I-two level}), the
Lewis-Riesenfeld phase (\ref{LRphase}) is now calculated as \cite{Lai}
\beqa
\label{phase-two level}
\alpha_{\pm} (t)
%&=& \frac{1}{\hbar}\int_0^t \Big{\langle}\phi_{\pm} (t')\Big|
%i \hbar \frac{\partial}{\partial t'} - H \Big| \phi_{\pm} (t') \Big{\rangle} dt',
%\\
= \pm \frac{1}{2} \int^t_0 [\Delta(t')  - 2 \tilde{\Omega}(t')] d t',
\eeqa
which gives
\beqa
\label{dotalpha}
\dot{\alpha}_{\pm}
%&=& \frac{1}{\hbar}\int_0^t \Big{\langle}\phi_{\pm} (t')\Big|
%i \hbar \frac{\partial}{\partial t'} - H \Big| \phi_{\pm} (t') \Big{\rangle} dt',
%\\
= \pm \frac{1}{2} \left[\Delta(t)  - 2 \tilde{\Omega}(t)\right],
\eeqa
where
$$
\tilde{\Omega} = (\Delta + \dot{\beta}) \cos^2{\left(\frac{\gamma}{2}\right)} + \frac{\Omega_R}{2}  \sin{\gamma}\cos{(\beta  - \varphi)}.
$$
Substituting Eq. (\ref{dotalpha}) into Eq. (\ref{inHa}),
we obtain
\beq
H(t) = \left(\frac{2 \tilde{\Omega} - \Delta}{\Omega_0}\right) I(t)  + i \hbar  \sum_{\pm} | \partial_t \phi_{\pm} (t) \ra \la \phi_{\pm} (t) |,
\eeq
which can be finally expressed as
\beqa
\label{inH-2level}
H (t)= \frac{\hbar}{2} \left(\begin{array}{cc} M & ~~N e^{i \beta}
\\
N^{\ast} e^{-i \beta} &   - M
\end{array}\right),
\eeqa
where
\begin{eqnarray*}
M= \Delta \cos^2{\gamma} + \Omega_R \sin{\gamma} \cos{\gamma} \cos{(\beta-\varphi)} - \dot{\beta}\sin^2{\gamma},
\\
N=[\Delta \cos{\gamma} + \Omega_R \sin{\gamma} \cos{(\beta- \varphi)} + \dot{\beta}\cos{\gamma}]\sin{\gamma} - i \dot{\gamma}.
\end{eqnarray*}
The Hamiltonian (\ref{inH-2level}) must be equivalent to the Hamiltonian (\ref{H-2level}),
so from $M= \Delta$ and $N e^{i \beta}= \Omega_R e^{i \varphi}$ \cite{Lai},
we get the following auxiliary equations,
\beqa
\label{gamma}
\dot{\gamma} &=& \Omega_R \sin{(\beta-\varphi)},
\eeqa
\beqa
\label{beta}
 (\Delta + \dot{\beta}) \sin{\gamma} &=& \Omega_R \cos{\gamma} \cos{( \beta- \varphi)}.
\eeqa
%
%can be obtained from $M= \Delta$ and $N e^{i \beta}= \Omega_R e^{i \varphi}$.
%In the particular case, $\sin {\gamma} = 0$, that is, $\gamma = n \pi$, we
%can further achieve
%\beqa
%\dot{\gamma} = \Omega_R ,  ~~~~~~  \beta- \varphi = 2 n \pi \pm \frac{\pi}{2} ,
%\eeqa
%or
%\beqa
%\dot{\gamma} = \Omega_R =0.
%\eeqa
%
%The latter ones result in $[H, I] =0$.

In general $H(t)$ does not commute with $I(t)$,
\beqa
\label{commute}
[H(t) , I (t)]/(\hbar^2\Omega_0) &=&  \nonumber  \hat{\sigma}_{+} \left(\Delta \sin{\gamma} e^{i \beta} - \Omega_R \cos{\gamma} e^{i \varphi} \right)  /2
\\ \nonumber
&-& \hat{\sigma}_{-} \left( \Delta \sin{\gamma} e^{-i \beta} - \Omega_R \cos{\gamma} e^{- i\varphi}  \right)  /2
\\
&+& \hat{\sigma}_0 i \Omega_R  \sin{\gamma} \sin{( \beta- \varphi)}.
\eeqa
$[H(0),I(0)]=0$ is satisfied, if
\beqa
\Delta(0) \sin{\gamma(0) } e^{i \beta(0) } - \Omega_R(0)  \cos{\gamma(0) } e^{i \varphi(0) }=0,
\\
\Delta(0)  \sin{\gamma(0) } e^{-i \beta(0) } - \Omega_R(0)  \cos{\gamma(0) } e^{- i\varphi(0) }=0,
\\
\Omega_R(0)   \sin{\gamma(0) } \sin{[ \beta(0) - \varphi(0) ]} = 0,
\eeqa
and there are similar equations for $t_f$.
%There are several mathematical solutions to the above equations.
For population inversion processes we are interested in processes starting and ending with
zero $\Omega_R$ and some finite detuning, so we impose
\beqa
\label{BC-1}
\Omega_R (0)  = 0, ~~\gamma (0) = n \pi,
\eeqa
so far with arbitrary $\beta(0)$ and $\varphi(0)$. In this case, $H(0)$ and $I(0)$ have
common eigenvectors, which are exactly the pure ground state $|1\ra$ and
the excited state $|2 \ra$.

Similarly, for $[H(t_f),I(t_f)]=0$,
\beqa
\Omega_R (t_f)  = 0, ~~\gamma (t_f) = n \pi,
\eeqa
with arbitrary $\beta(t_f)$ and $\varphi(t_f)$. Again $H(t_f)$ and $I(t_f)$
share common eigenstates, which are the ground and the excited state.

Substituting the above boundary conditions into Eqs. (\ref{gamma})
and (\ref{beta}), we further obtain
\beqa
\label{BC-3}
\dot{\gamma} (0)=0, ~~ \dot{\gamma} (t_f)=0,
\eeqa
whereas $\dot{\beta} (0)$ and $\dot{\beta} (t_f)$ will determine the value of the initial and final detunings.

%In addition,
%
%\beqa
%\left \{\begin{array}{cc}
%\label{BC-1}
%\beta (0)  - \varphi (0) =  2 n \pi,
%\\
%\tan{\gamma (0)} = \Omega_R (0)/ \Delta(0),
%\end{array}\right.
%\eeqa
%
%
%\beqa
%\left \{\begin{array}{cc}
%\beta (0)  -   \varphi (0)= (2n+1) \pi,
%\\
%\tan{\gamma (0)} = - \Omega_R (0)/ \Delta(0).
%\end{array}\right.
%\eeqa
%

We are now ready to set
some ansatz for $\beta$ and $\gamma$ using appropriate boundary conditions.
Once the functions $\beta$ and $\gamma$
are fixed, we can construct $\Omega_R$ and $\Delta$ and thus the Hamiltonian $H(t)$
with a given $\varphi$. In the following subsection,
we will give some examples to show how the invariant-based engineering
method works with different boundary conditions.
\subsection{Example}
\begin{figure}[]
\begin{center}
\scalebox{0.45}[0.45]{\includegraphics{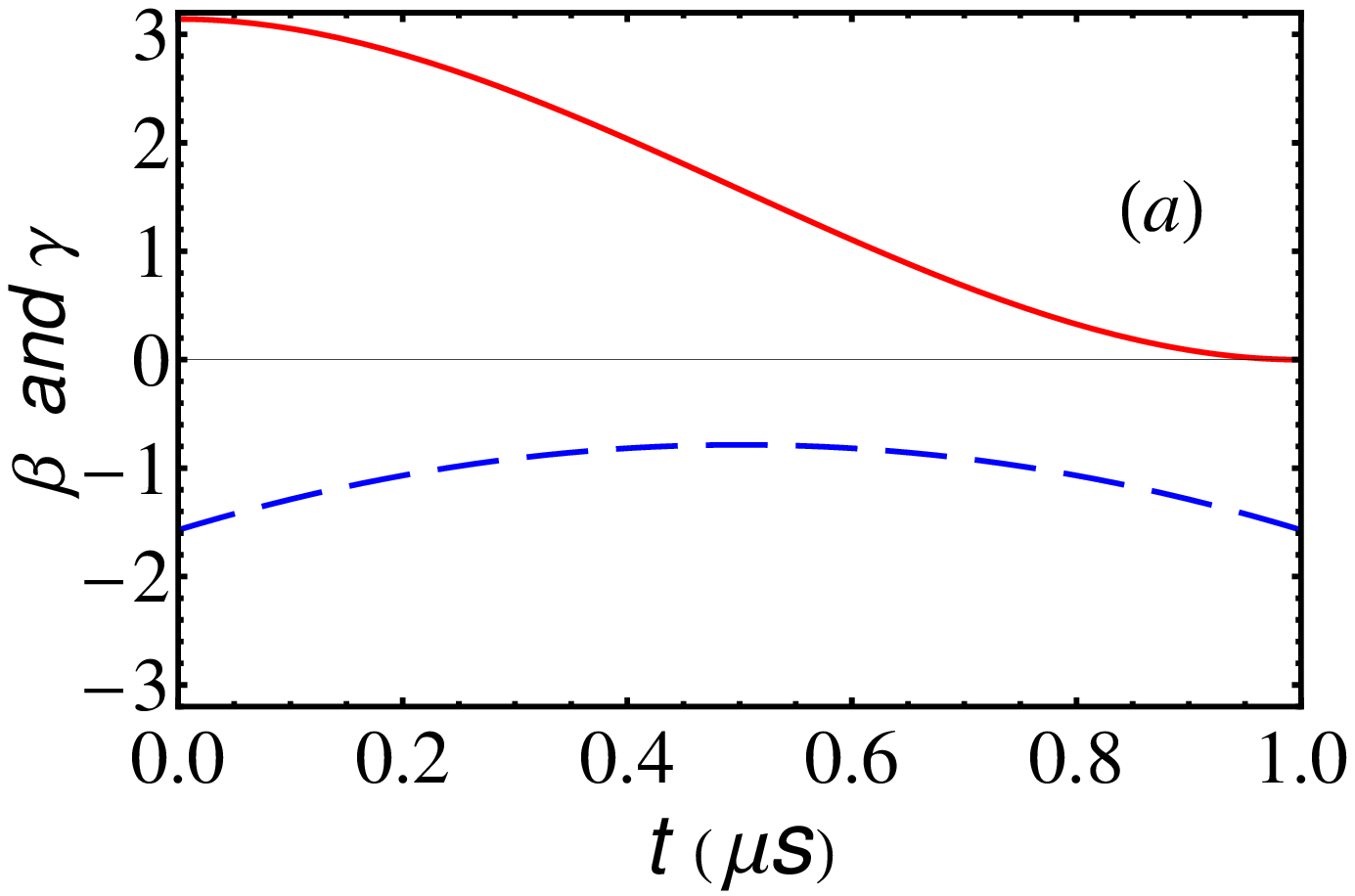}}
\scalebox{0.45}[0.45]{\includegraphics{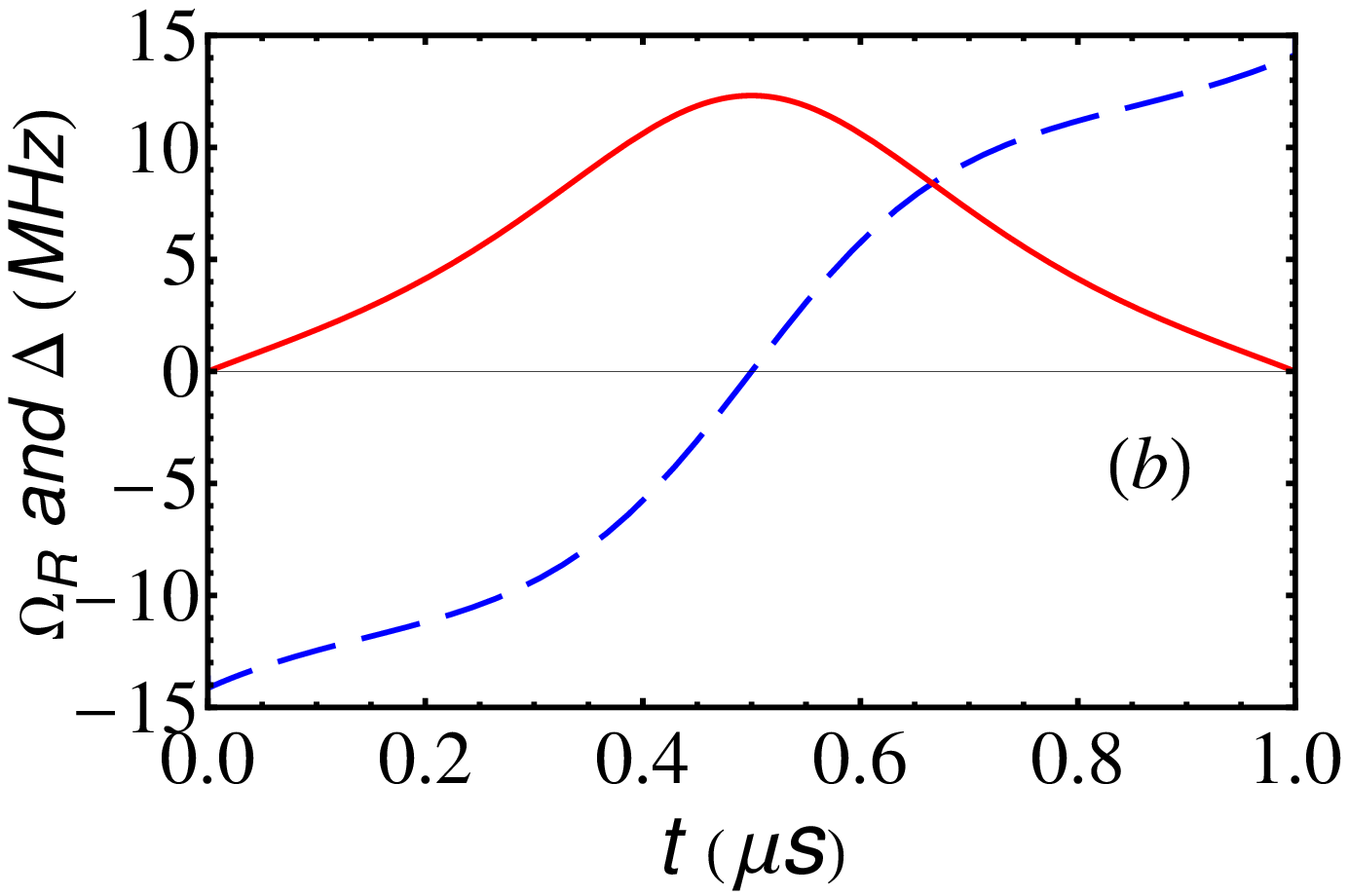}}
\scalebox{0.45}[0.45]{\includegraphics{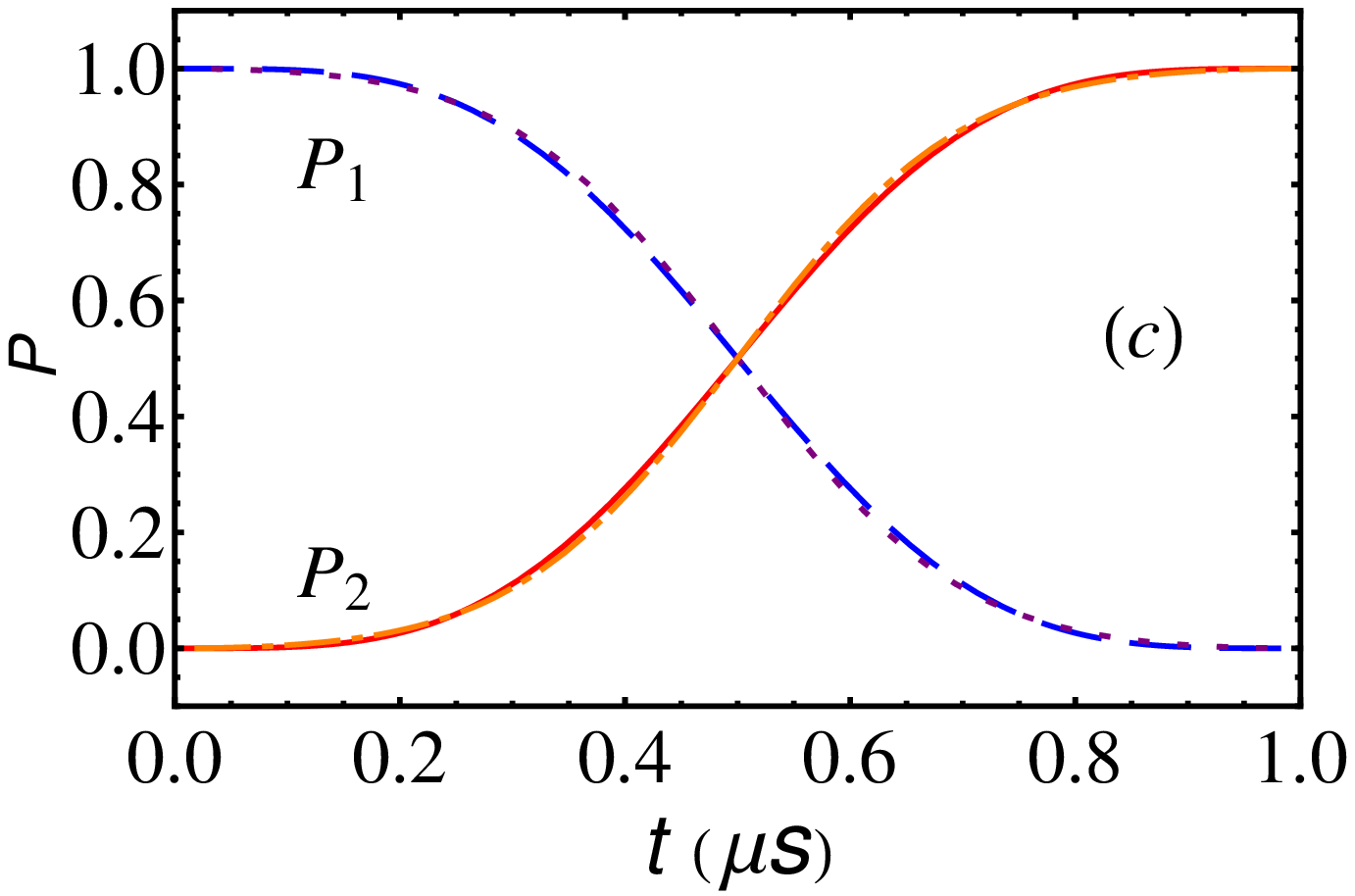}}
 \caption{(Color online). (a) Examples of polynomial ansatzs for $\gamma (t) = \sum^3_{j=0} a_j t^j$ (solid red line) and $\beta (t) = \sum^3_{j=0} b_j t^j$ (dashed blue line). (b) The corresponding functions of $\Omega_R$ (solid red line) and $\Delta$ (dashed blue line) determined by Eqs. (\ref{omega}) and (\ref{delta}). (c) Time evolution of the populations of levels $1$ and $2$: $P_1$ (dashed blue line),
$P_2$ (solid red line), and adiabatic approximations $P^{ad}_1$ (dotted purple line)  and $P^{ad}_2$ (dash-dotted orange line), see Eq. (\ref{pad}), hardly distinguishable from the former.
$t_f = 1 \mu s$. } \label{Fig.1}
\end{center}
\end{figure}
We shall apply the previous results to design fast population transfer protocols.
For simplicity, we assume $\varphi =0$ and consider the, yet unknown, Hamiltonian $H(t)$,
\beqa
\label{H-2level-simplified}
H (t)= \frac{\hbar}{2} \left(\begin{array}{cc} \Delta &\Omega_{R}
\\
\Omega_{R} & -\Delta
\end{array}\right),
\eeqa
where $\Omega_R$ and $\Delta$ are determined, from (\ref{gamma}) and (\ref{beta})  by
\beqa
\label{omega}
\Omega_R &=& \dot{\gamma}/\sin \beta, ~~~~
\\
\label{delta}
\Delta &=& \Omega_R \cot{\gamma} \cos{\beta}-\dot{\beta}.
\eeqa
We suppose that this Hamiltonian drives the state from  $|1 \rangle$
to $|2 \rangle$, up to the phase factor, along the invariant eigenvector
$|\phi_{+} (t) \ra$. To this end,
we set the boundary conditions $\Omega(0) = \Omega(t_f)= 0$ and
\beqa
\label{BC-gamma}
\gamma (0)= \pi, ~~ \dot{\gamma} (0)= 0, %~~\beta (0)=0,  %~~\dot{\beta} (0)=0,
\\
\gamma (t_f)=0, ~~ \dot{\gamma} (t_f)=0. %~ \beta (t_f)=0, %~~ \dot{\beta} (t_f)=0,
\eeqa
As mentioned before, we have freedom to
choose the values of $\beta(0)$ and $\beta(t_f)$. According to Eq. (\ref{omega}), it is useful  to keep $\beta$ close to $(n+1/2) \pi$, so as to minimize $\Omega_R$ along the path, whereas the derivatives fix the initial and final detunings, see Eq. (\ref{delta}), which should have here opposite signs.
Moreover they should not be too large to keep $\beta$ close to the chosen reference $\beta$ value, and to avoid $\beta=0$ at some intermediate time
and thus an infinite $\Omega_R$.
Considering all these physical constraints, we impose
\beqa
\beta (0) = -\pi/2,  ~~ \dot{\beta} (0) = 3 \pi/(2 t_f),
\\
\label{BC-beta}
 \beta (t_f) = -\pi/2,  ~~ \dot{\beta}  (t_f) = -3 \pi/(2 t_f),
\eeqa
where the negative sign of $\beta$, see Eq. (\ref{omega})
and Fig. \ref{Fig.1} (a),  keeps $\Omega_R$
positive, as $\dot{\gamma}$ becomes negative.

To interpolate at intermediate times we assume a polynomial ansatz.
Fig. \ref{Fig.1} (a) shows
$\gamma (t) = \sum^3_{j=0} a_j t^j$ and $\beta (t) = \sum^3_{j=0} b_j t^j$,
where the coefficients are obtained by solving the equations set by the boundary conditions.
The time-dependent $\Omega_R$ and $\Delta$
calculated from Eqs. (\ref{omega}) and (\ref{delta}) are shown in Fig. \ref{Fig.1} (b).
Once we have specified $H(t)$ in (\ref{H-2level-simplified}), we solve the dynamics numerically by a
Runge-Kutta method with adaptive step, see the population inversion in
Fig. \ref{Fig.1} (c) for levels 1 and 2. We have also compared the bare state populations
$P_1$ and $P_2$ with the populations of the instantaneous eigenstates of $H(t)$,
$P^{ad}_1$ and $P^{ad}_2$. Their agreement shows that the designed protocol
is in fact an adiabatic passage for the specified final time $t_f$.

\begin{figure}[]
\begin{center}
\scalebox{0.45}[0.45]{\includegraphics{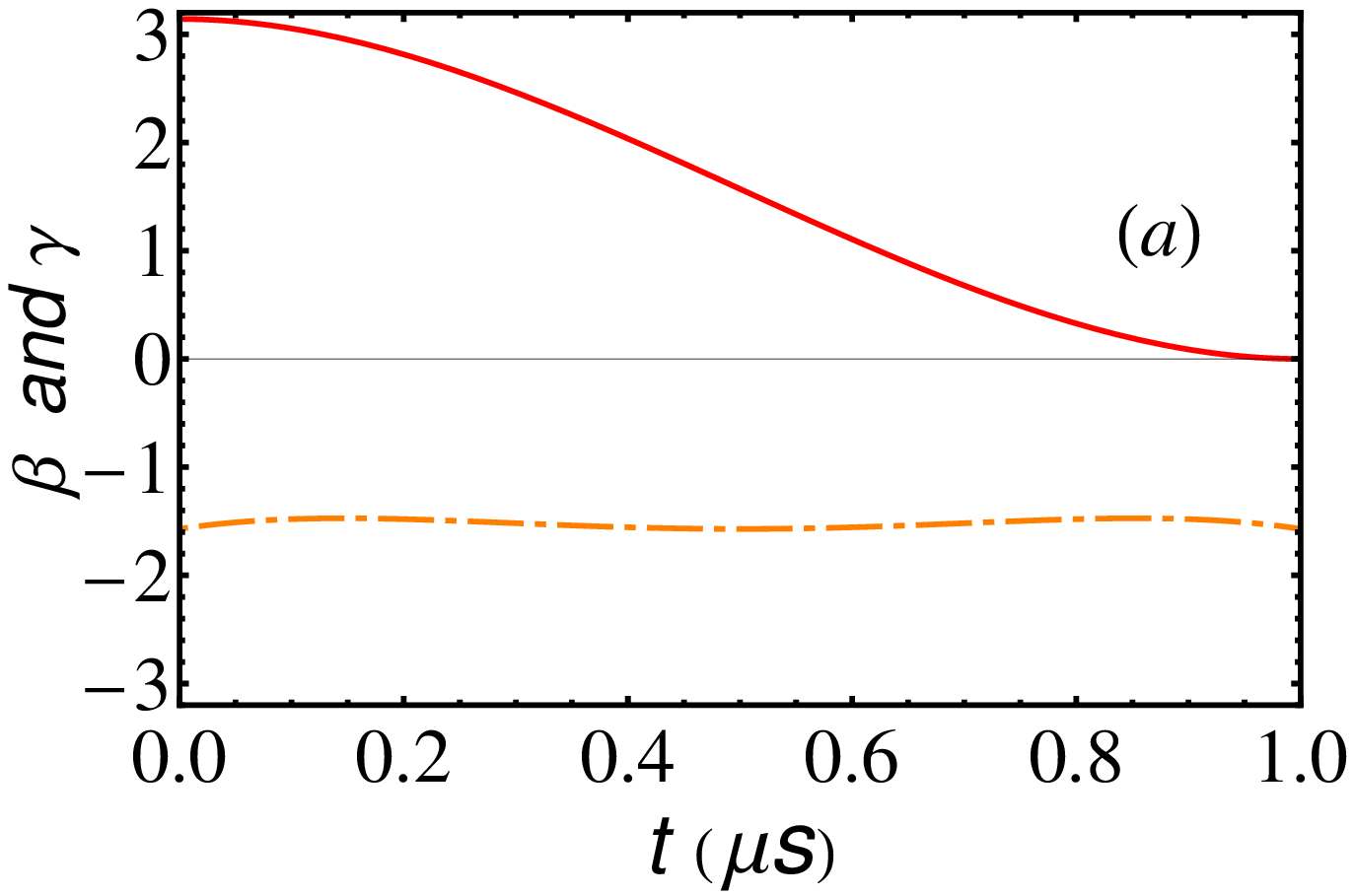}}
\scalebox{0.45}[0.45]{\includegraphics{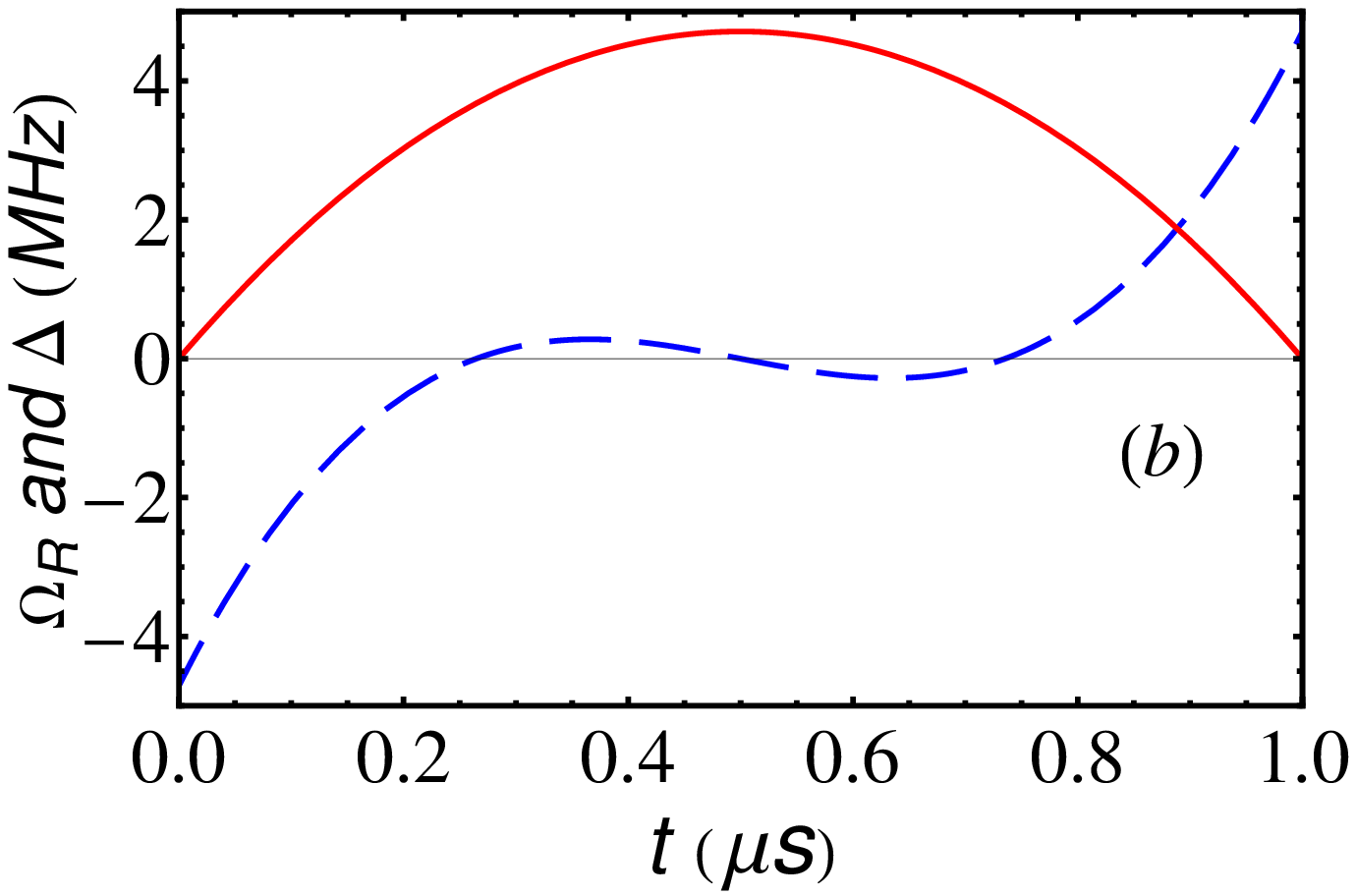}}
\scalebox{0.45}[0.45]{\includegraphics{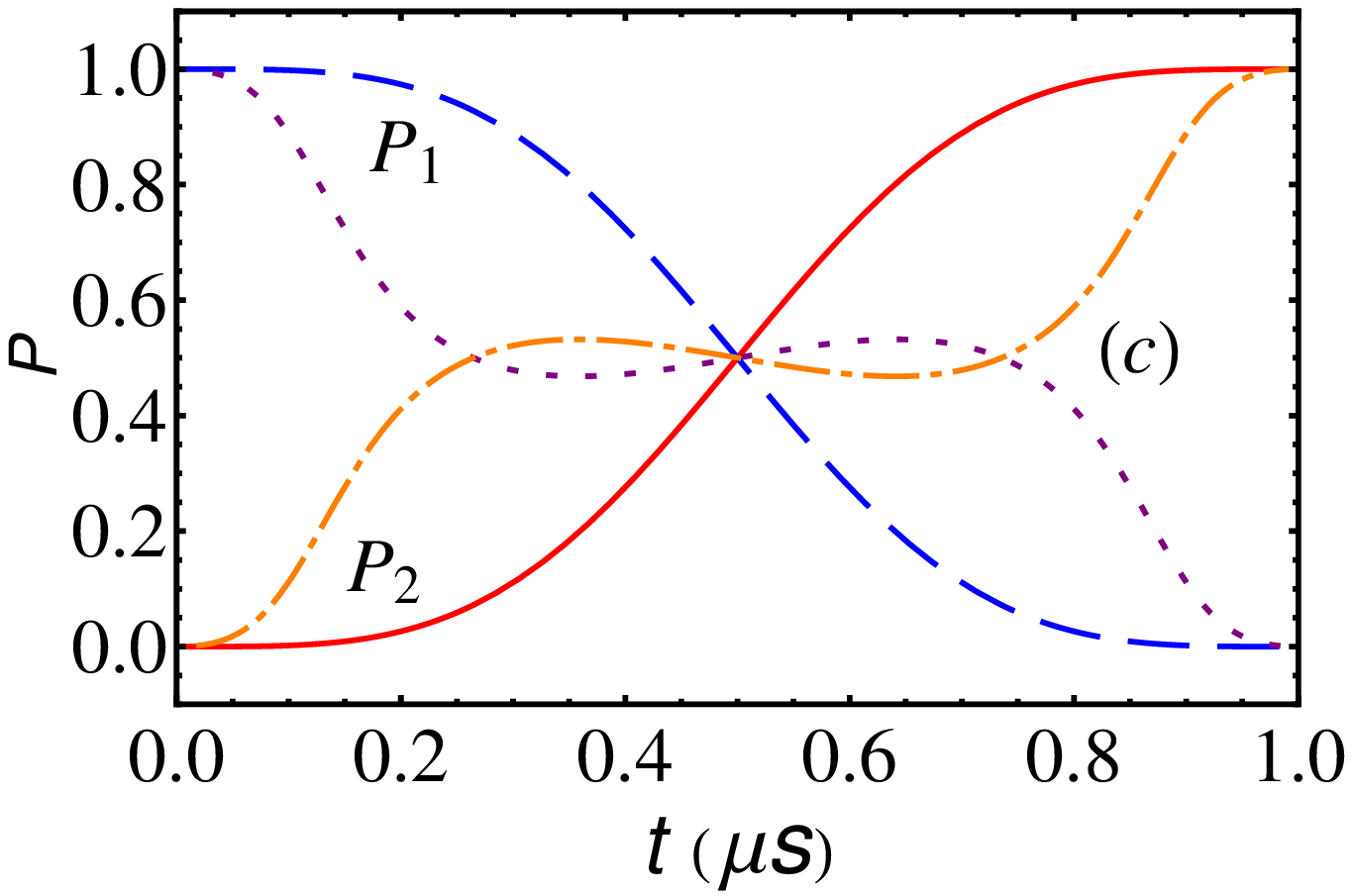}}
 \caption{(Color online). (a) Examples of polynomial ansatzs for $\gamma (t) = \sum^3_{j=0} a_j t^j$ (solid red line) and $\beta (t) = \sum^4_{j=0} b_j t^j$ (dashed blue line). (b) The corresponding functions of $\Omega_R$ (solid red line) and $\Delta$ (dashed blue line) determined by Eqs. (\ref{omega}) and (\ref{delta}). (c) Time evolution of the populations $P_1$ (dashed blue line) and $P_2$ (solid red line); adiabatic approximations
$P^{ad}_1$ (dotted purple line) and $P^{ad}_2$ (dash-dotted orange line) for comparison.
 $t_f = 1 \mu s$. } \label{Fig.2}
\end{center}
\end{figure}

We may impose additional conditions at an intermediate time, for example,
to keep $\beta$ closer to $-\pi/2$,
\beqa
\beta (0) = -\pi/2,  ~~ \beta (t_f) = -\pi/2, ~~ \beta (t_f/2) = -\pi/2,
\\
\label{BC-beta-2}
\dot{\beta} (0) = \pi / (2t_f),    ~~ \dot{\beta} (t_f) = - \pi/(2 t_f),
\eeqa
where we have also diminished the detuning.
This new set of conditions requires a higher other polynomial,
$\beta (t) = \sum^4_{j=0} b_j t^j$.
Fig. \ref{Fig.2} shows the results, to be compared with those of Fig. \ref{Fig.1}.
Note that in Fig. \ref{Fig.2} (b) Rabi frequency and detunings are
smaller than in Fig. \ref{Fig.1} (b), so smaller energies are involved.
Now the dynamical evolution is not adiabatic, see Fig. \ref{Fig.1} (c).
%We have thus managed to realize fast population transfer
%by inverse engineering.
%the invariant-based engineering method, which to our knowledge
%hasn't been discussed yet \cite{Lai}.
The method can be further complemented by optimizing the trajectory
with respect to different physical cost functions or constraints \cite{Li},
for example, setting bounds for $\Omega_R$, whose square is proportional
to the laser intensity.
This will be discussed elsewhere.
%
%state transfer begins with the pure bare state $|1\ra$ and ends up with $|2 \ra$ along the mode $|\phi_{-}(t) \ra $,
%the parameter $\gamma$ in the eignestate $| \phi_{-} (t) \ra $ should change from $0$ to $\pi$.
%Since $I(t)$ and $H(t)$ have the common eigenstates at initial and final times, the mixing angle $\theta$ in the state $| n_{-} (t)\ra $
%should also vary from $\pi$ to $0$ correspondingly. Basically, $\beta(0) + \varphi(0) = (2 n+1) \pi$ and
%$\beta(t_f) + \varphi(t_f) = (2 n+1) \pi$ should be also satisfied in the meanwhile.
%Based on Eq. (\ref{gamma}), the Rabi frequency can be calculated from
%$\Omega_R = \dot{\gamma}/ \sin(\varphi + \beta )$, so that in order to avoid the infinite value of Rabi frequency in the duration,
%function $\beta + \varphi$ is forbidden to go through the singularity $2n \pi$ or $(2 n+1) \pi$.
%Taking into account these aforementioned requirements, we can express finally the boundary conditions in the following way,
%
%\beqa
%\label{BC-example}
%\gamma(0) = \pi + \arctan{\left[\frac{\Omega_R (0)}{\Delta (0)}\right]}, ~~~~ \beta(0) = \pi -\varphi(0),
%\\
%\gamma(t_f) =\arctan{\left[\frac{\Omega_R (t_f)}{\Delta (t_f)}\right]}, ~~~~~ \beta(t_f) = \pi -\varphi(t_f),
%\eeqa
%
%
%
\subsection{Invariant-based method in transitionless tracking algorithm language}
In order to reexamine the invariant-based inverse engineering approach in the language of Berry's transtionless tracking algorithm,
we take $|\phi_{\pm}(t)\ra$ as $|n_{0\pm}(t)\ra$, and let $ \alpha_{\pm} (t) = \xi_{\pm} (t)$,
so that Eq. (\ref{general H_0}) gives
\beqa
\nonumber
H_0 (t) &=& \frac{\Delta \cos \gamma+ \Omega_R \sin \gamma \cos{(\beta-\varphi)}}{\Omega_0} I(t)
\\
 &=&  \frac{2 \tilde{\Omega} - \Delta - 2 \dot{\beta} \cos^{2}{(\gamma/2)}}{\Omega_0} I(t),
\eeqa
and the Hamiltonian $H_1(t)$ in Eq. (\ref{Berry Hamiltonian}) is
\beq
H_1 (t) = \frac{\hbar}{2} \left(\begin{array}{cc} - \dot{\beta}\sin^2{\gamma} & (- i \dot{\gamma} + \frac{\dot{\beta}}{2} \sin{2\gamma}) e^{i \beta}
\\
(i \dot{\gamma} +  \frac{\dot{\beta}}{2}  \sin{2\gamma}) e^{- i \beta} &   \dot{\beta}\sin^2{\gamma}
\end{array}\right).
\eeq
Using the boundary conditions (\ref{BC-1})-(\ref{BC-3}), $H_1(t)$ vanishes
at $t=0$ and $t=t_f$.
\subsection{Transitionless tracking algorithm}
Let us now apply Berry's transitionless tracking algorithm
taking the Hamiltonian (\ref{H-2level}) as reference Hamiltonian $H_0 (t)$ \cite{Ch10b},
\beq
\label{2level}
H_{0} (t) = \frac{\hbar}{2} \left(\begin{array}{cc} \Delta & \Omega_{R} e^{i \varphi}
\\
\Omega_{R} e^{-i \varphi} &   -\Delta
\end{array}\right).
\eeq
The driving Hamiltonian (\ref{Berry Hamiltonian}) becomes in this case
\beq
\label{Berry H_1}
H_1 (t) = \frac{\hbar}{2} \left(\begin{array}{cc} - \dot{\varphi}\sin^2{\theta} & (- i \dot{\theta} + \frac{\dot{\varphi}}{2} \sin{2\theta}) e^{i \varphi}
\\
( i \dot{\theta} + \frac{\dot{\varphi}}{2} \sin{2\theta}) e^{- i \varphi}
 &   \dot{\varphi}\sin^2{\theta}
\end{array}\right).
\eeq
%with $\Omega_a\equiv\dot{\theta}=(\Omega_R \dot{\Delta}-\dot{\Omega}_R \Delta)/\Omega^2$,
%$\Omega_b \equiv - \dot{\varphi} \sin\theta \cos\theta = -\dot{\varphi}(\Omega_R \Delta/\Omega^2)$ and
%$\Delta_b \equiv - \dot{\varphi} \sin^2{\theta} = - \dot{\varphi} (\Omega_R/\Omega)^2$.
For $\varphi =0$,
%
%\beq
%H_{0} (t) = \frac{\hbar}{2} \left(\begin{array}{cc} \Delta & \Omega_{R}
%\\
%\Omega_{R}  &   -\Delta
%\end{array}\right),
%\eeq
this reduces to
\beqa
H_1 (t)=  \frac{\hbar}{2} \left(\begin{array}{cc} 0 &  i \Omega_a
\\
- i \Omega_a &  0
\end{array}\right),
\eeqa
where $\Omega_a\equiv\dot{\theta}=(\Omega_R \dot{\Delta}-\dot{\Omega}_R \Delta)/\Omega^2$.
In \cite{Ch10b}, this was used to speed up an Allen-Eberly scheme for $H_0(t)$ and achieve  fast population transfer.
To see the physical meaning and realizability of the
method we must go back to the Schr\"{o}dinger picture:
For $H(t)=H_0(t)+H_1(t)$ this implies using two lasers with
the same frequency, orthogonal polarization, and  time-dependent intensities but different intensity shapes \cite{Ch10b}.
The alternative is to drive the system with $H_1(t)$ only, without $H_0(t)$.
In the Schr\"odinger picture this amounts to act with one laser and
to perform level shift engineering to modulate the transition frequency so as to
leave $\Delta=0$ in the interaction picture \cite{Ch10b}.
Note that these complications (an extra laser or the need for level-shift engineering)
do not arise in the results obtained in the previous subsection.
\subsection{Relation to invariants}
Let us now reinterpret the previous (standard) Berry's transitionless tracking algorithm in terms of invariant theory.
In the language of Lewis-Riesefeld invariant theory, we can construct an invariant as
$I(t)= \sum_{\pm} |n_{0 \pm} (t) \ra \lambda_{\pm} \la n_{0 \pm}(t)|$, where   $\lambda_{\pm} = E^{(0)}_{\pm} (0) =\pm \hbar \Omega_0/2$,
with matrix form
\beqa
I (t)= \frac{\hbar}{2} \Omega_0 \left(\begin{array}{cc} \cos{\theta} & \sin{\theta} e^{i \varphi}
\\ \sin{\theta} e^{- i \varphi} & - \cos{\theta}
\end{array}\right),
\eeqa
and $I (0) = H_0 (0)$.
Since $H_0 (t) = \sum_{\pm}  |n_{0 \pm} (t) \ra E^{(0)}_{\pm}  (t)\la n_{ 0 \pm}(t)| $ with instantaneous eigenvalues
$E^{(0)}_{\pm}  (t) = \pm \hbar \Omega(t)/2$, we have
$H_0 (t) = [\Omega(t)/\Omega_0] I(t) $.
Using $ \dot{\xi}_{\pm} = - E^{(0)}_{\pm} (t) /\hbar + i \la  n_{0 \pm} (t)| \partial_t n_{0 \pm}(t) \ra $ and
$\la  n_{0 \pm}(t) | \partial_t n_{0 \pm} (t)\ra = \pm i \dot{\varphi} \cos^2{(\theta/2)}$, then letting $| \phi_{\pm}(t) \ra  = |n_{0 \pm} (t)\ra$ and $\alpha_{\pm} (t)
= \xi_{\pm} (t)$, we may write $H(t)$ from Eq. (\ref{inHa}).
%as
%
%\beqa
%H (t) = \left[\frac{\Omega + 2 \dot{\varphi} \cos^2{\left(\frac{\theta}{2}\right)}}{\Omega_0}\right] I(t) +
%H_1(t) - \frac{2 \dot{\varphi} \cos^2{\left(\frac{\theta}{2}\right)}}{\Omega_0} I(t).
%\eeqa
%
Canceling terms this gives exactly the Hamiltonian $H_0(t)+H_1(t)$
in the previous subsection.
$I(t)$ does not commute with $H(t)$ in general, but, when the boundary conditions
\beqa
\theta  (0) = \pi,~~  \dot{\theta} (0)=0,
\\
\theta (t_f) = 0, ~~ \dot{\theta} (t_f)=0,
\eeqa
are satisfied, $H_1 (t)$ will vanish at initial and final times.
%, as discussed in Sec. \ref{SecIVA}.
%it is guaranteed that $I (0)$ [$I(t_f)$] and $H(0)$ [$H(t_f)$] have the common eigenstates at $t=0$ and $t=t_f$, %respectively.
%
%
%
%
%
%
\section{Conclusion}
In previous publications we applied and compared two methods to speed-up adiabatic processes through  non-adiabatic shortcuts: Berry's transitionless
tracking algorithm and the invariant-based inverse engineering approach. Their differences were emphasized,
in particular in time-dependent harmonic oscillators \cite{Muga10}
or transport of particles by a moving trap \cite{transport}. The message here is quite different, even opposite: we point out now that in fact both approaches share a common ground of concepts and structure. There is, however, no contradiction. It is indeed possible to interpret a particular inverse engineering operation using either the language of transitionless-tracking or invariants approaches, and  consider them to be potentially equivalent. The explanation of the differences found is the large freedom to design different Hamiltonians for a given speed-up goal.
In other words, the different results
are not fundamental but due to the particular choices that have been made to resolve that freedom in specific implementations.
The choice of method from this point becomes thus, in part,
a matter of taste, but there are also elements
that make one or the other approach more natural or convenient.
For example, systems with Hamiltonians that admit known structures
for the invariants are easy to approach with the invariant-based method.
This includes transport, expansions, rotations \cite{Lohe}, or, as shown here,
discrete level systems.
The tracking algorithm can be applied in many systems where the
invariants are unknown.
In summary this work provides a significant step towards a deeper understanding of shortcut-to-adiabaticity methods that will help to choose the most adequate approach in atomic transport, quantum gates,
and generally atomic manipulation and control applications.
\section*{Acknowledgments}
We thank M. Berry for commenting on the manuscript.
This work was supported by Basque Government (Grant No. IT 472-10), Ministerio de Ciencia e Innovaci\'on (Grant No. FIS2009-12773-C02-01), Juan de la Cierva Programme,
the National Natural Science Foundation of China (Grant No. 60806041), and the Shanghai Leading Academic Discipline
Program (Grant No. S30105). E. T. acknowledges support from the Basque Government (Grant No. BFI08.151).

\end{document}